\documentclass[sigconf]{acmart}

\AtBeginDocument{%
  \providecommand\BibTeX{{%
    \normalfont B\kern-0.5em{\scshape i\kern-0.25em b}\kern-0.8em\TeX}}}

\usepackage{natbib}
\usepackage{multirow}
\usepackage{multicol}

\makeatletter
\renewcommand{\NAT@separator}{\NAT@sep\nolinebreak}
\makeatother

\begin{document}

\title{Deduplicating and Ranking Solution Programs for Suggesting Reference Solutions}

\author{Atsushi Shirafuji}
\affiliation{
    \institution{Graduate Department of Computer\\and Information Systems\\University of Aizu}
    \city{Aizu-Wakamatsu}
    \country{Japan}
}
\email{m5261161@u-aizu.ac.jp}

\author{Yutaka Watanobe}
\affiliation{
    \institution{Department of Computer Science\\and Engineering\\University of Aizu}
    \city{Aizu-Wakamatsu}
    \country{Japan}
}
\email{yutaka@u-aizu.ac.jp}

\renewcommand{\shortauthors}{A. Shirafuji and Y. Watanobe}

\begin{abstract}
Referring to solution programs written by other users is helpful for learners in programming education.
However, current online judge systems just list all solution programs submitted by users for references, and the programs are sorted based on the submission date and time, execution time, or user rating, ignoring to what extent the programs can be helpful to be referenced.
In addition, users struggle to refer to a variety of solution approaches since there are too many duplicated and near-duplicated programs. 

To motivate learners to refer to various solutions to learn better solution approaches, in this paper, we propose an approach to deduplicate and rank common solution programs in each programming problem.
Inspired by the nature that the many-duplicated program adopts a more common approach and can be a general reference, we remove the near-duplicated solution programs and rank the unique programs based on the duplicate count.

The experiments on the solution programs submitted to a real-world online judge system demonstrate that the number of programs is reduced by 60.20\%, whereas the baseline only reduces by 29.59\% after the deduplication, meaning that users only need to refer to 39.80\% of programs on average.
Furthermore, our analysis shows that top-10 ranked programs cover 29.95\% of programs on average, indicating that users can grasp 29.95\% of solution approaches by referring to only 10 programs.
The proposed approach shows the potential of reducing the learners' burden of referring to too many solutions and motivating them to learn a variety of solution approaches.
\end{abstract}



\maketitle

\section{Introduction}
\label{sec:introduction}

Programming novices learn correct and better approaches to solving programming problems by referring to various solution programs written by expert programmers or instructors.
Among Japanese programmers, \emph{shaky\={o}-style learning} (originally meaning the sutra copying, hand-copying of Buddhist sutras), which is manually typing the reference source code, is considered one of the ways of learning programming for novices.

Using e-learning systems in programming education is becoming more popular as online judge systems can automatically and immediately evaluate users' submitted programs~\cite{wasik2018oj, watanobe2022aoj}.
As some online judge systems publicly show the submitted solution programs, such as Aizu Online Judge (AOJ)\footnote{\url{https://onlinejudge.u-aizu.ac.jp/}.} and AtCoder\footnote{\url{https://atcoder.jp/}.}, users can refer to other users' solution approaches.

Referring to solution programs written by other users is important for learners in programming education.
However, current online judge systems just list all solution programs submitted by users for references, and the programs are ranked based on several criteria, such as the submission date and time, execution time or memory, and user ratings, which ignore to what extent the program can be a reference.
The effective (e.g., faster and less computational cost) programs can be unreadable since experienced programmers may write complicated programs for effectiveness, leveraging the advanced algorithms or data structures.
Similarly, higher-rating users can write more complicated programs for effectiveness.
Therefore, the performance of programs and the user ratings cannot be good metrics for sorting solution programs for novice programmers to use as references.

Moreover, users struggle to refer to a variety of solution approaches since there are too many duplicated and near-duplicated programs.
For example, a set of introductory programming problems, \texttt{ITP1}, provided on AOJ, each programming problem has 7,430 solution programs on average in Python.
It is clear that novice programmers struggle to find the standard solution approaches from the thousands of solution programs when they want to refer to better solution approaches after solving or giving up on solving the problem.

To motivate the learners to refer to various reference solutions to learn better solution approaches, in this paper, we propose an approach to deduplicate and rank solution programs submitted by users in each programming problem.
The approach is inspired by the nature that \textbf{a many-duplicated program adopts a more common approach and can be a reference}.
We remove near-duplicated programs and rank the unique (deduplicated) programs based on the duplicate count.

The advantages of the proposed approach are as follows.
\begin{itemize}
    \item Deduplication makes it easy to browse a variety of solution approaches.
    \item The suggested programs are ensured to be correct.
    \item Through the code formatter, programs with consistent readability are stably suggested.
\end{itemize}

The quantitative evaluation of the proposed approach shows that it significantly reduces the number of programs after deduplication, with an average reduction of 60.20\%.
This means that users only need to refer to 39.80\% of the programs on average.
Furthermore, the top-10 ranked programs cover an average of 29.95\% of the solution programs, indicating that users can grasp nearly 30\% of the solution approaches by referring to only 10 programs.
Although the top-ranked programs do not necessarily indicate that they are the most helpful programs, these results demonstrate the effectiveness of the proposed approach in reducing the learners' burden of referring to too many solutions and improving the coverage of solution programs.

The contributions of this work are as follows.
\begin{itemize}
    \item The proposal of an approach to deduplicate and rank solution programs in programming education, which helps learners find common reference solutions.
    \item The evaluation of the proposed approach using a real-world dataset of introductory programming problems shows its effectiveness in reducing the number of programs and improving the coverage of solution programs.
    \item Identifying limitations and future directions for the proposed approach, including the code format and the need for a sufficient number of solution programs for reliable suggestions.
\end{itemize}

\section{Related Work}
\label{sec:related-work}

\begin{figure*}[h]
    \centerline{\includegraphics[width=\linewidth]{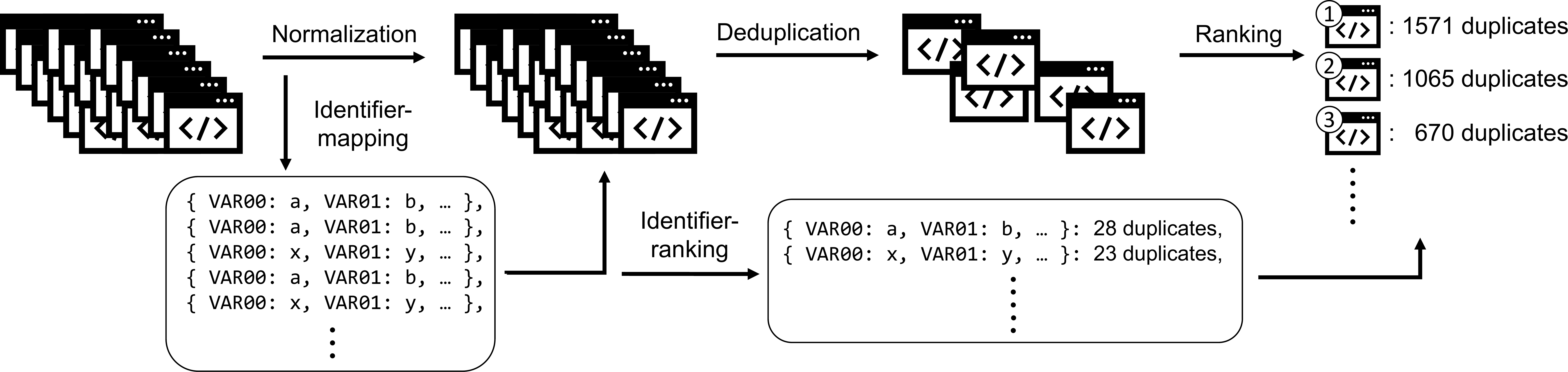}}
    \caption{Illustration of the proposed approach.}
    \label{fig:approach}
\end{figure*}

There are several prior works that are related to our work on deduplicating and ranking solution programs for suggesting reference solutions.
We divide the related work into three categories: source code deduplication, online judge systems, and code-to-code generation.

\subsection{Source Code Deduplication}
\label{sec:related-work:deduplication}

Source code deduplication is a task to identify and remove duplicate or near-duplicate source code, which is semantically or syntactically similar source code.
This task is usually performed in creating large-scale source code datasets~\cite{puri2021codenet, kocetkov2023stack} or training machine learning models on source code~\cite{fried2023incoder, allal2023santacoder} to avoid redundancy, overfitting, and memorization~\cite{allamanis2019duplicates, lee2022deduplicating, kandpal2022deduplicating}.
Not only based on the exact match but also the code clone detection task can be applied for deduplication.
Code clone detection algorithms typically consider the syntactic or semantic similarity between code snippets and apply various techniques such as tokenization and abstract syntax tree (AST) analysis.
Several tools have been developed for this task, such as CloneDR~\cite{baxter1998clonedr}, CCFinder~\cite{kamiya2002ccfinder}, and DECKARD~\cite{Jiang2007deckard}.
While these techniques are useful for identifying duplicate or near-duplicate programs, they are not directly applicable to our task of suggesting reference solutions to programming problems because our goal is to restore the complete working program among the near-duplicate programs.

\subsection{Online Judge Systems}
\label{sec:related-work:ojs}

Another related area of research is online judge systems, which are platforms that allow users to submit solution programs to a set of programming problems and automatically evaluate the submitted programs~\cite{wasik2018oj}.
Many online judge systems, such as AOJ and AtCoder, make the submitted solutions publicly available, allowing other users to refer to them for educational and reference purposes.
This allows learners to study and learn from existing solution programs and explore different approaches to solving the same problem.
However, the existing online judge systems do not provide effective mechanisms for deduplicating and ranking the solution programs.
They often sort the programs based on criteria such as submission date and time, execution time, or user ratings\footnote{AtCoder's Rating System, \url{https://www.dropbox.com/s/ixci4amralioaif/rating.pdf}.}\footnote{Ranking Criteria - Aizu Online Judge, \url{https://judge.u-aizu.ac.jp/onlinejudge/rating_note.jsp}.}, which may not reflect the quality or popularity of the programs.
Our work aims to address this limitation and enhance the usefulness of these solution programs by proposing an approach to deduplicate and rank the solution programs based on their popularity.

\subsection{Code-to-Code Generation}
\label{sec:related-work:code-to-code}

Code-to-code generation is a task to generate code based on a given input code.
This task is closely related to our work as we aim to generate reference solution programs from the user-submitted solution programs.
It is important to clearly distinguish our work from code repair or error correction tasks~\cite{rahman2021repair, matsumoto2021repair}, generating correct programs from incorrect programs, whereas our work only focuses on correct programs in both input and output.
There has been some research on code-to-code generation tasks, such as code recommendation and code refactoring~\cite{luan2019recommend, silavong2022recommend}.
These tasks involve generating code similar to the given input code but improved in some way, such as being more efficient, readable, or maintainable.
For example, the work by Madaan~et~al.~\cite{madaan2023improving} demonstrated the use of large language models (i.e., Codex~\cite{chen2021codex} and CodeGen~\cite{nijkamp2022codegen}) for improving the execution time of given programs.
Similarly, Madaan~et~al.~\cite{madaan2023self-refine} proposed improving code readability by iteratively refining code generation.
While these approaches are similar to ours in modifying existing code to improve it, they focus on improving individual programs rather than deduplicating and ranking solution programs for suggesting reference solutions.
In our work, we perform deduplication and ranking based on the duplicate count, leveraging the AST to capture the structural similarity of the solution programs.
Furthermore, our approach does not require any input programs from the user, as we aim to suggest existing solution programs rather than generating new ones.

In summary, while there has been prior work on source code deduplication, online judge systems, and code-to-code generation, our work focuses on the specific problem of deduplicating and ranking solution programs in programming education.
We propose an approach that considers the semantic equivalence and popularity of the programs to provide helpful reference solutions for learners.

\section{Methodology}
\label{sec:method:method}

The proposed approach is primarily inspired by the nature that \textbf{the more duplicated programs are more common and can be references for learners}.
To suggest the more common and helpful reference solution programs to users, we propose an approach to remove duplicated programs and re-rank the user-submitted solution programs based on their popularity (i.e., duplicated count).
The proposed approach is illustrated in Figure~\ref{fig:approach}.
The approach can be divided into three phases: \textit{normalization}, \textit{deduplication}, and \textit{ranking}.

\subsection{Normalization}
\label{sec:method:normalization}

\begin{figure*}[h]
    \centerline{\includegraphics[width=\linewidth]{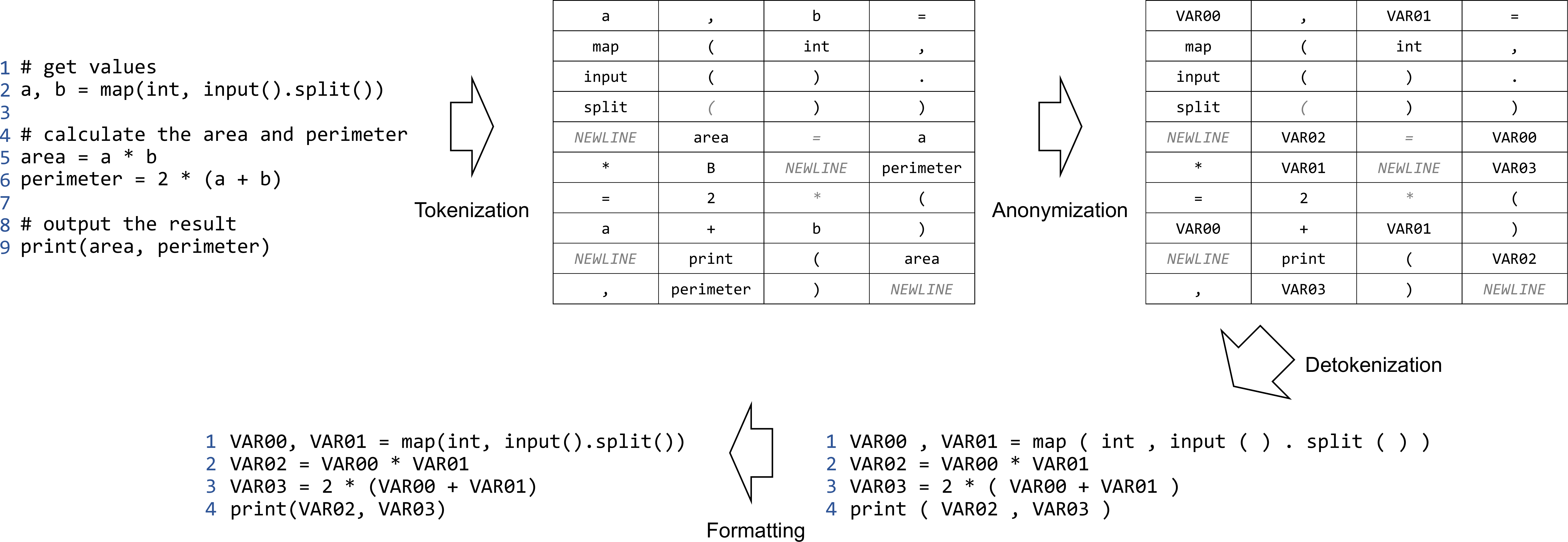}}
    \caption{Illustration of normalization, which is divided into four phases: tokenization, anonymization, detokenization, and formatting.}
    \label{fig:normalization}
\end{figure*}

In the normalization phase, to better capture the syntactic equality of the programs in the deduplication phase, we normalize the superficial representations of the programs.
In more detail, we ignore the white lines, white spaces, comments, docstrings, and identifier names (e.g., variable, function, and class names).
The normalization can be further divided into four phases: \textit{tokenization}, \textit{anonymization}, \textit{detokenization}, and \textit{formatting}.
Figure~\ref{fig:normalization} illustrates the normalization.

Although several prior works used AST to capture the more abstract structure of the programs~\cite{yamaguchi2012ast, yoshizawa2018ast}, using source code rather than AST is more suitable in this work since we need to reconstruct the source code from AST to suggest the complete working source code for users.
However, we also use the AST in the anonymization phase to decide which name should be anonymized (e.g., functions defined in the standard library should not be anonymized, whereas the user-defined functions should be anonymized).

\subsubsection{Tokenization}
\label{sec:method:normalization:tokenization}

For tokenization, we develop a lightweight tokenizer using the standard \texttt{tokenize}\footnote{\url{https://docs.python.org/3/library/tokenize.html}.} library in Python to remove comments and docstrings completely.
In general, docstrings cannot be removed in tokenization because a docstring can be considered a string literal not assigned to any variables.
Similarly, comments using string literals cannot be removed.
Therefore, we develop a tokenizer to support removing the comments and docstrings by removing the string literals not assigned to any variables.

\subsubsection{Anonymization}
\label{sec:method:normalization:anonymization}

To normalize the identifier names (e.g., variable, function, and class names), we employ anonymization.
The anonymization is inspired by the normalization phase of prior works~\cite{bhoopchand2016pycodesuggest,terada2021completion}, leveraging the \texttt{ast}\footnote{\url{https://docs.python.org/3/library/ast.html}.} and \texttt{astor}\footnote{\url{https://github.com/berkerpeksag/astor}.} libraries.

Although the normalized programs only use the anonymized variable names, we also collect the original ones.
We create a map of anonymized to original identifier names (hereinafter, \textit{identifier map}) to restore the original variable names from the anonymized ones, called \textit{identifier-mapping}.
This identifier map will be used in the final ranking phase in Section~\ref{sec:method:ranking} to restore the most popular variable names from the duplicated programs.

\subsubsection{Detokenization \& Formatting}
\label{sec:method:normalization:detokenization-formatting}

Detokenization is to restore the source code from the normalized tokens.
This reconstruction is a process to back into human-readable source code after normalization, performed at the token level.

After the detokenization, we apply a code formatter to format the detokenized source code.
In the detokenization phase, concatenating each token with a space can make the source code unreadable because of excessive spaces.
Although this does not pose a semantic error, suggesting source code that adopts an uncommon format could be harmful to learners.
We employ the \texttt{yapf}\footnote{\url{https://github.com/google/yapf}.} library as the code formatter.

\subsection{Deduplication}
\label{sec:method:deduplication}

After the normalization, we deduplicate the normalized programs based on the exact match, which considers the programs to be the same if the normalized programs are exactly the same.

The duplicated programs are removed in this phase.
We refer to the programs after the deduplication as \textit{unique programs}.
For each unique program, we count the number of duplicated programs and collect the identifier maps for the next ranking phase.

\subsection{Ranking}
\label{sec:method:ranking}

The ranking phase has two steps to make a list of common reference programs: (1) program ranking and (2) identifier ranking.
Program ranking is to make a list of common programs by sorting the unique programs for each problem based on the duplicate count of the program.
Identifier ranking is to make a list of common identifier names by sorting the identifier maps for each unique program based on the duplicate count of the identifier map.
Both rankings consider the popularity based on duplicate count.

In the identifier ranking, we count the duplicated identifier names using the identifier maps.
To keep the consistency of using the identifiers and to avoid altering the original code, the identifier map is only considered a duplicate if all the identifiers are the same.
For instance, identifier maps such as $\{ \text{VAR01} = x, \text{VAR02} = y \}$, and $\{ \text{VAR01} = x, \text{VAR02} = s \}$ are considered different due to the difference in the variable $\text{VAR02}$.
If we count duplicates for individual variables, we may select $\{ \text{VAR01} = s, \text{VAR02} = s \}$, which does not appear in any solution programs and breaks the behavior.
We apply this method as it eliminates the risk of selecting inconsistent identifier names and guarantees to suggest a complete working program.

\section{Experiment}
\label{sec:experiment}

We use introductory programming problems and apply the proposed methodology to demonstrate how it contributes to suggesting better solution programs with less user burden in referring to too many programs.

\subsection{Dataset}
\label{sec:experiment:dataset}

For the target programming problems, we use a set of introductory programming problems provided on AOJ~\cite{watanobe2004aoj}.
AOJ is one of the popular online judge systems where users submit programs to solve given programming problems~\cite{wasik2018oj}.
AOJ currently stores approximately 8 million programs submitted by 100 thousand users for 3 thousand programming problems~\cite{watanobe2022aoj}.

We target a course named \textit{Introduction to Programming I} (\texttt{ITP1})\footnote{\url{https://onlinejudge.u-aizu.ac.jp/courses/lesson/2/ITP1/all}.}.
The course has 44 introductory programming problems, ranging from standard input/output to class definition.

The submitted programs include not only correct programs but also wrong or incomplete programs.
Each submission has a \textit{verdict}, an evaluation result of executing hidden test cases on the judge system.
For instance, \textit{Accepted} (AC) indicates a correct program, \textit{Runtime Error} (RE) indicates an incorrect program failed in execution, and \textit{Wrong Answer} (WA) also indicates an incorrect program that the output result is wrong.
There are more verdicts\footnote{\url{https://onlinejudge.u-aizu.ac.jp/judges_replies}.}.
However, we only use the submissions judged as AC in this work.
In addition, we target Python~3 programs.

Moreover, we only use public programs, which users set their code policy to make the submitted programs publicly available, following the Terms of Use\footnote{\url{https://onlinejudge.u-aizu.ac.jp/term_of_use}.} of AOJ.
The public programs can be used for research or educational purposes and are collected in popular datasets, such as CodeNet~\cite{puri2021codenet} and CodeContests~\cite{li2022alphacode}.

\begin{table}
\small
\begin{center}
\begin{minipage}{\linewidth}
    \caption{Dataset statistics of 44 problems. \textit{Solutions} indicates correct programs, \textit{Available} indicates publicly available correct programs, and \textit{Outliers} indicates programs failed in normalization. We use \textit{Valid} in the following sections.}
    \label{table:dataset}
    \begin{tabular*}{\linewidth}
    {lrrrrr}
        \toprule
         & \#Submissions & \#Solutions & \#Available & \#Outliers & \#Valid \\
        \midrule
        Total & 868,143 & 326,933 & 150,497 & 14 & 150,483 \\
        Mean & 19,731 & 7,430 & 3,420 & 0 & 3,420 \\
        Std & 13,462 & 5,204 & 2,191 & 1 & 2,191 \\
        \bottomrule
    \end{tabular*}
\end{minipage}
\end{center}
\end{table}

Table~\ref{table:dataset} shows the number of programs used in this work.
\texttt{ITP1} had 868,143 submissions at the time.
After filtering only AC programs, the number of solution programs is 326,933.
Still, it contains private programs and programs that are not archived yet.
After filtering only public and archived programs, the number of available solution programs is 150,497.
Fourteen programs are further filtered out as outliers because they fail in conversion into AST.
Finally, the number of valid solution programs is 150,483.
On average, each programming problem has 3,420 solution programs.

\begin{table}
\small
\begin{center}
\begin{minipage}{0.95\linewidth}
    \caption{Decrease of the number of programs after deduplication on 44 problems and their average. Baseline is based on the exact match deduplication without normalization. \#Unique indicates the number of unique programs after deduplication, and ratio indicates the ratio of unique programs out of solution programs. \#Unique and ratio are smaller the better.}
    \label{table:decrease}
    \begin{tabular*}{\linewidth}
    {lrrrrr}
        \toprule
        Problem & \#Solutions & \multicolumn{2}{c}{Baseline} & \multicolumn{2}{c}{\textbf{Ours}} \\
        & & \#Unique & Ratio & \#Unique & Ratio \\
        \midrule
        \texttt{ITP1\_1\_A} & 10,699 & 380 & 3.55\% & 81 & 0.76\% \\ 
        \texttt{ITP1\_1\_B} & 9,044 & 1,954 & 21.61\% & 350 & 3.87\% \\ 
        \texttt{ITP1\_1\_C} & 8,004 & 4,649 & 58.08\% & 1,446 & 18.07\% \\ 
        \texttt{ITP1\_1\_D} & 6,782 & 5,154 & 76.00\% & 2,223 & 32.78\% \\ 
        \texttt{ITP1\_2\_A} & 5,954 & 4,000 & 67.18\% & 1,011 & 16.98\% \\ 
        \texttt{ITP1\_2\_B} & 5,240 & 3,104 & 59.24\% & 661 & 12.61\% \\ 
        \texttt{ITP1\_2\_C} & 5,434 & 4,254 & 78.28\% & 2,035 & 37.45\% \\ 
        \texttt{ITP1\_2\_D} & 4,768 & 4,130 & 86.62\% & 1,937 & 40.63\% \\ 
        \texttt{ITP1\_3\_A} & 5,191 & 1,541 & 29.69\% & 234 & 4.51\% \\ 
        \texttt{ITP1\_3\_B} & 4,784 & 3,905 & 81.63\% & 1,797 & 37.56\% \\ 
        \texttt{ITP1\_3\_C} & 4,727 & 3,983 & 84.26\% & 2,331 & 49.31\% \\ 
        \texttt{ITP1\_3\_D} & 4,207 & 3,604 & 85.67\% & 1,185 & 28.17\% \\ 
        \texttt{ITP1\_4\_A} & 4,246 & 3,470 & 81.72\% & 1,866 & 43.95\% \\ 
        \texttt{ITP1\_4\_B} & 3,927 & 3,511 & 89.41\% & 2,338 & 59.54\% \\ 
        \texttt{ITP1\_4\_C} & 3,999 & 3,462 & 86.57\% & 2,009 & 50.24\% \\ 
        \texttt{ITP1\_4\_D} & 4,036 & 3,199 & 79.26\% & 1,340 & 33.20\% \\ 
        \texttt{ITP1\_5\_A} & 3,770 & 3,195 & 84.75\% & 1,576 & 41.80\% \\ 
        \texttt{ITP1\_5\_B} & 3,326 & 3,037 & 91.31\% & 1,929 & 58.00\% \\ 
        \texttt{ITP1\_5\_C} & 3,204 & 2,923 & 91.23\% & 1,979 & 61.77\% \\ 
        \texttt{ITP1\_5\_D} & 2,432 & 1,973 & 81.13\% & 1,478 & 60.77\% \\ 
        \texttt{ITP1\_6\_A} & 3,379 & 2,754 & 81.50\% & 1,495 & 44.24\% \\ 
        \texttt{ITP1\_6\_B} & 3,070 & 2,707 & 88.18\% & 2,190 & 71.34\% \\ 
        \texttt{ITP1\_6\_C} & 2,551 & 2,227 & 87.30\% & 1,847 & 72.40\% \\ 
        \texttt{ITP1\_6\_D} & 2,250 & 2,014 & 89.51\% & 1,487 & 66.09\% \\ 
        \texttt{ITP1\_7\_A} & 2,379 & 2,133 & 89.66\% & 1,658 & 69.69\% \\ 
        \texttt{ITP1\_7\_B} & 3,633 & 3,224 & 88.74\% & 2,360 & 64.96\% \\ 
        \texttt{ITP1\_7\_C} & 2,032 & 1,822 & 89.67\% & 1,574 & 77.46\% \\ 
        \texttt{ITP1\_7\_D} & 1,831 & 1,597 & 87.22\% & 1,331 & 72.69\% \\ 
        \texttt{ITP1\_8\_A} & 2,166 & 1,252 & 57.80\% & 570 & 26.32\% \\ 
        \texttt{ITP1\_8\_B} & 2,218 & 1,995 & 89.95\% & 1,089 & 49.10\% \\ 
        \texttt{ITP1\_8\_C} & 2,082 & 1,753 & 84.20\% & 1,435 & 68.92\% \\ 
        \texttt{ITP1\_8\_D} & 1,986 & 1,677 & 84.44\% & 857 & 43.15\% \\ 
        \texttt{ITP1\_9\_A} & 1,920 & 1,699 & 88.49\% & 1,269 & 66.09\% \\ 
        \texttt{ITP1\_9\_B} & 1,720 & 1,522 & 88.49\% & 854 & 49.65\% \\ 
        \texttt{ITP1\_9\_C} & 1,696 & 1,553 & 91.57\% & 1,089 & 64.21\% \\ 
        \texttt{ITP1\_9\_D} & 1,514 & 1,376 & 90.89\% & 1,196 & 79.00\% \\ 
        \texttt{ITP1\_10\_A} & 1,773 & 1,542 & 86.97\% & 862 & 48.62\% \\ 
        \texttt{ITP1\_10\_B} & 1,543 & 1,418 & 91.90\% & 1,306 & 84.64\% \\ 
        \texttt{ITP1\_10\_C} & 1,524 & 1,404 & 92.13\% & 1,229 & 80.64\% \\ 
        \texttt{ITP1\_10\_D} & 1,417 & 1,277 & 90.12\% & 1,165 & 82.22\% \\ 
        \texttt{ITP1\_11\_A} & 1,292 & 1,173 & 90.79\% & 1,025 & 79.33\% \\ 
        \texttt{ITP1\_11\_B} & 904 & 835 & 92.37\% & 750 & 82.96\% \\ 
        \texttt{ITP1\_11\_C} & 938 & 815 & 86.89\% & 741 & 79.00\% \\ 
        \texttt{ITP1\_11\_D} & 891 & 759 & 85.19\% & 683 & 76.66\% \\ 
        \midrule
        Total & 150,483 & 105,956 & --- & 59,868 & --- \\
        Mean & 3,420 & 2,408 & 70.41\% & 1,361 & \textbf{39.80\%} \\
        \bottomrule
    \end{tabular*}
\end{minipage}
\end{center}
\end{table}

\subsection{Results}
\label{sec:experiment:results}

\subsubsection{Deduplication}
\label{sec:experiment:results:deduplication}

Table~\ref{table:decrease} shows the decrease in the number of programs after deduplication, compared with the baseline approach.
The baseline skips the normalization phase and directly deduplicates based on the exact match of the raw source code.
Compared with the baseline, our proposed approach can remove more near-duplicated programs, resulting in fewer unique programs.
On average, the proposed approach reduces the number of programs from 3,420 to 1,361, which reduces by 60.20\%, whereas the baseline approach only reduces from 3,420 to 2,408, which reduces by 29.59\%.
Our approach reduces the number of programs by 43.48\% compared with the baseline.
This suggests that users can grasp more solution approaches by referring to fewer programs.

\begin{figure}
    \centerline{\includegraphics[width=\linewidth]{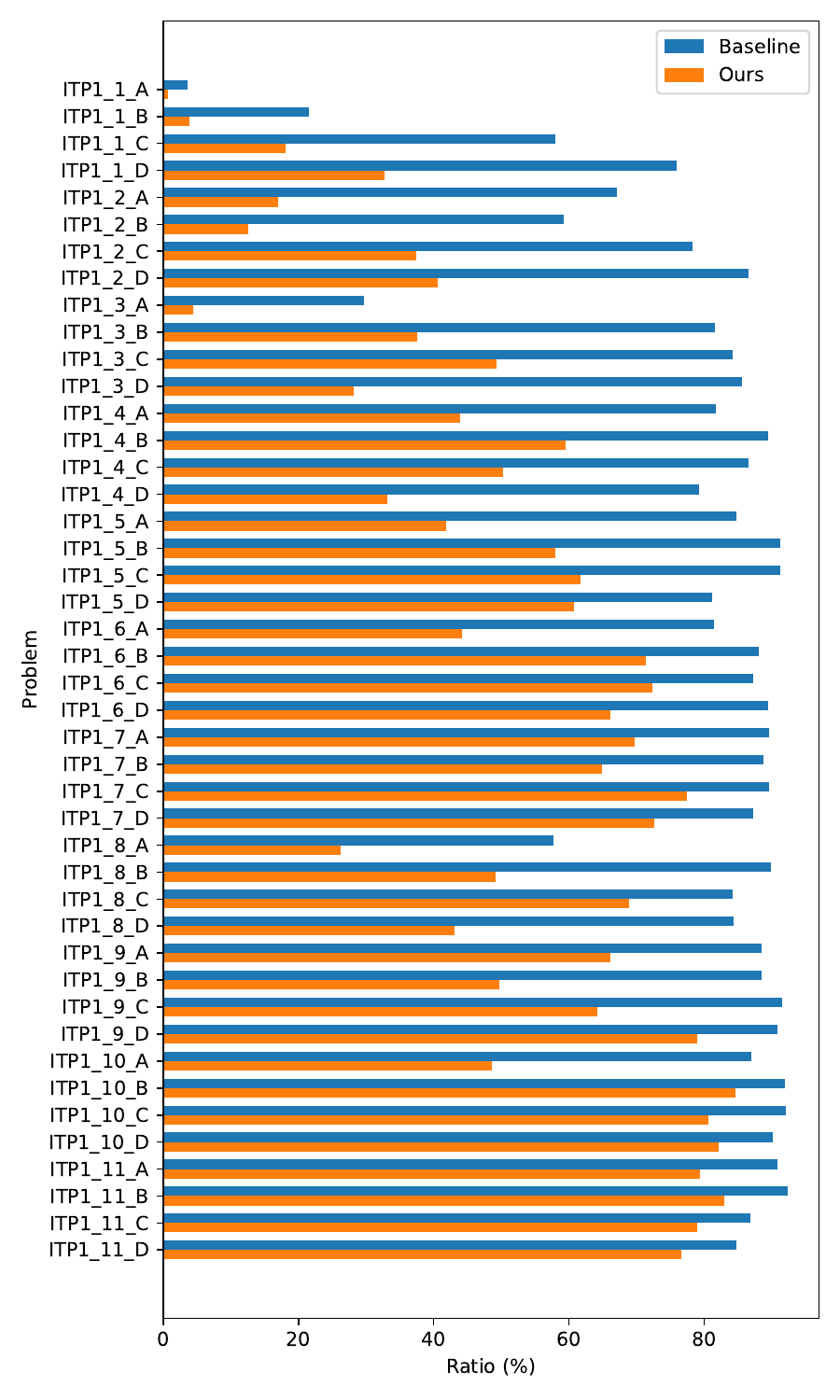}}
    \caption{The comparison between the baseline and our proposed approaches of the ratio of unique programs out of solution programs. The smaller the better.}
    \label{fig:problem-ratio}
\end{figure}

As shown in Figure~\ref{fig:problem-ratio}, we observe that the initial problems have a higher rate of reduction (i.e., a lower ratio of unique programs) in the number of programs.
This is because, in \texttt{ITP1}, the easier problems are located at the beginning and tend to have simple solution approaches.
Conversely, the reduction is significantly low for more challenging problems like \texttt{ITP1\_11}, where a class definition is required to solve.
We find that the proposed approach is highly effective on easier programming problems, whereas it is less effective on more complex programming problems with more diversity in solution approaches.

\subsubsection{Ranking}
\label{sec:experiment:results:ranking}

\begin{figure}
    \centerline{\includegraphics[width=\linewidth]{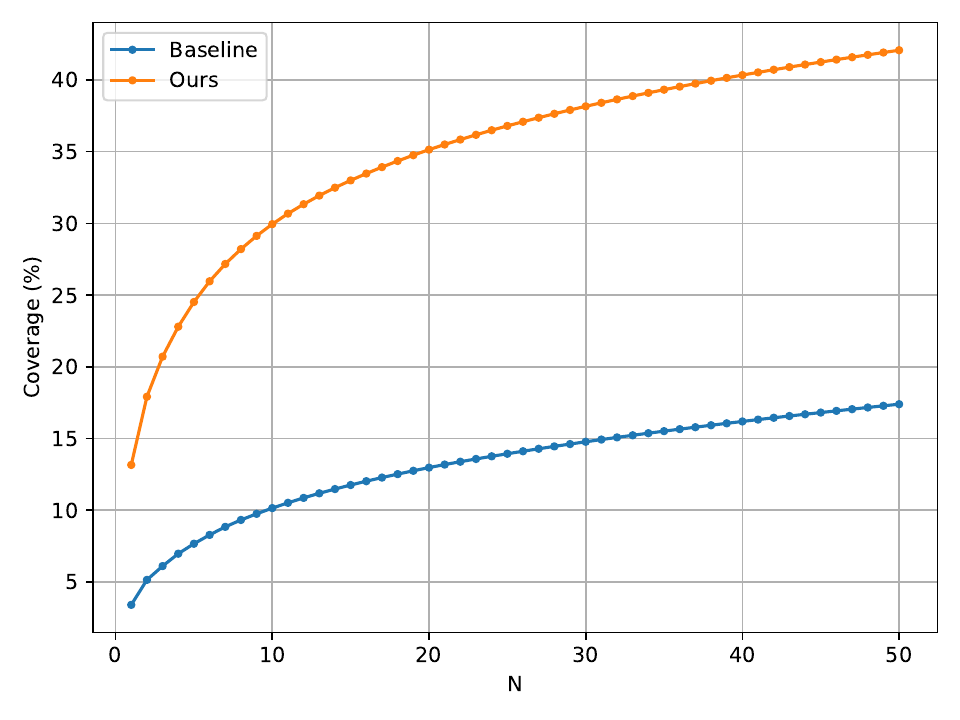}}
    \caption{Average solution coverage on top-$n$ ranked programs on 44 problems. The higher the better.}
    \label{fig:coverage}
\end{figure}

Figure~\ref{fig:coverage} shows the coverage of solution programs by top-$n$ ranked unique programs, comparing with the baseline approach based on the exact match deduplication without normalization.
The average coverage of 13.16\% for the top-1 program indicates that the users can grasp 13.16\% of the solution programs by referring to that one program.
Similarly, the coverage of 29.95\% for the top-10 programs indicates that the users can grasp approximately 30\% of the solution programs by referring to only 10 programs.
This shows the contribution of the proposed approach for users to refer to fewer programs to grasp solution approaches used to solve a programming problem.

However, the increase in coverage with an increase in $N$ does not necessarily show a good result.
Although a higher coverage indicates that users can grasp more solution approaches, an increase in $N$ implies an increase in the user burden of referring to $N$ solutions.
Therefore, this is a trade-off relationship.
In future work, we expect higher coverage with smaller $N$.

\subsubsection{Qualitative Evaluation}
\label{sec:experiment:results:qualitative}

As the qualitative evaluation, we manually evaluate the deduplicated (suggested) programs.
In each problem, the system suggests top-5 programs with top-3 identifiers for each program.

\begin{figure}
    \centerline{\includegraphics[width=\linewidth]{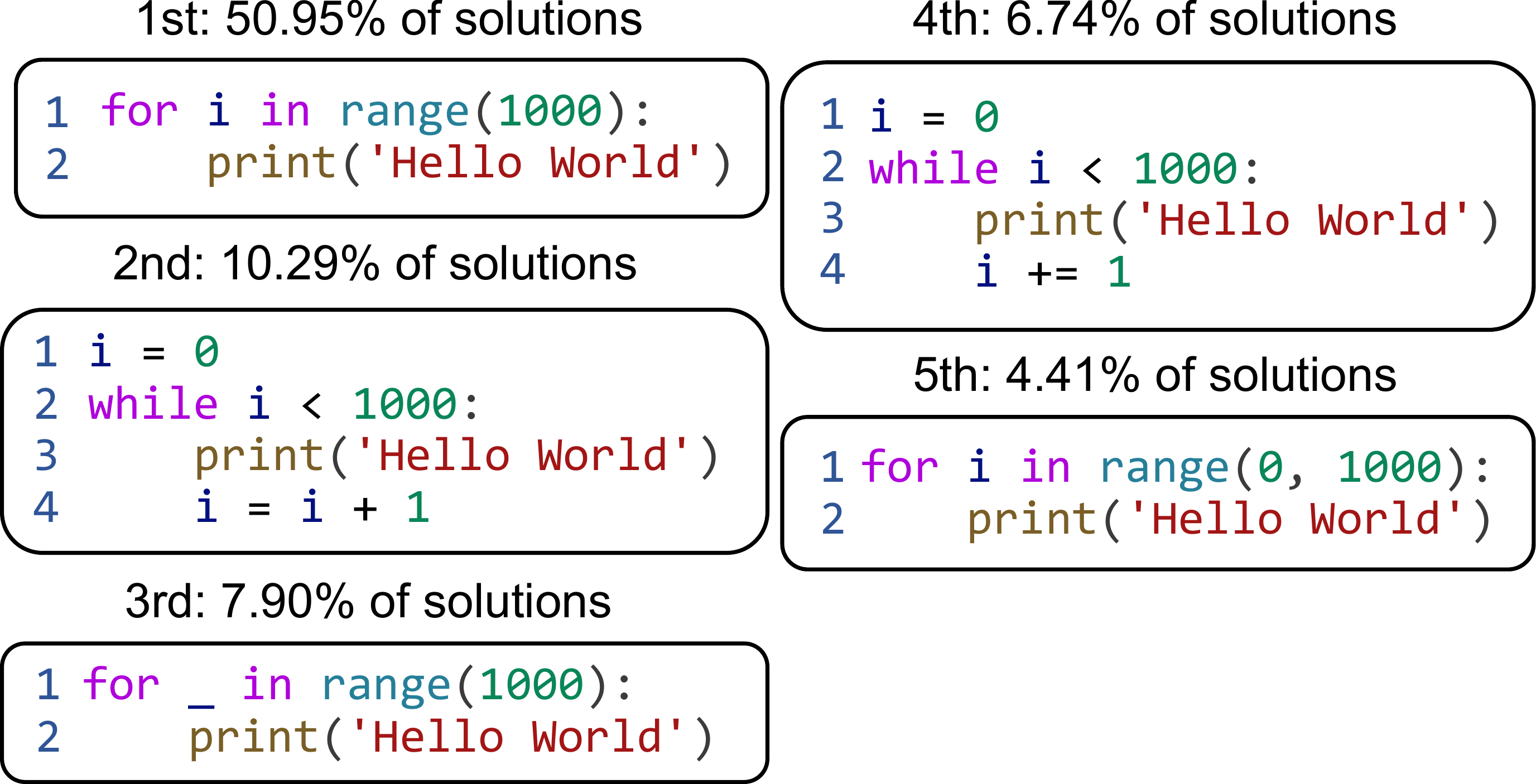}}
    \caption{Examples of top-5 programs for \texttt{ITP1\_3\_A}.}
    \label{fig:examples}
\end{figure}

\paragraph{Helpfulness:}
Figure~\ref{fig:examples} shows an example of top-5 programs for \texttt{ITP1\_3\_A}, which is to output ``Hello World'' 1,000 times.
Overall, the suggested programs are mostly readable and can be helpful references.
In the example, users can instantly grasp that the programming problem can be solved using a for-loop or while-loop with a print function.

However, some programs can still be considered near-duplicates as some use the same approaches with a slight difference.
In addition, we find that the top-ranked programs do not necessarily indicate that they are the most helpful programs.
In the example, the most helpful solution is not the 1st-ranked program, as the 3rd-ranked program defines the unused loop variable $i$ as an anonymous variable \texttt{\_} (underscore).
Similarly, the 4th-ranked program is better than the 2nd-ranked program as it uses an augmented assignment operator \texttt{+=} to update the variable $i$.

While there is no guarantee that the best solution approaches will be suggested, it can provide common and helpful solution programs with a certain degree of reliability.

\paragraph{Code Readability:}
For the readability of the suggested programs, we find that the normalization phase negatively impacted the code format, although we used a code formatter in the final phase.
This is primarily due to removing tokens such as comments, white lines, and white spaces in the normalization.
Normalization is necessary to better detect the near-duplicated programs.
Although the code formatter added white lines and spaces if required, we find that the original programs are more readable than the suggested programs.
Furthermore, removing comments and docstrings leaves the code without helpful explanations of functions and variables, making it less comprehensible for other users.
Although it would be beneficial to incorporate comments back into the code, directly using user-written comments can be challenging.
Further improvement of the approach can be delegated to the task of generating natural language comments for the given program.

\section{Limitation}
\label{sec:limitation}

Although the proposed approach has various advantages described in Section~\ref{sec:introduction}, it still has the following limitations.

First, it requires a certain amount of solution programs already submitted by users for reliable suggestions.
The number of user solutions influences the ranking based on the number of duplicates.
Therefore, it cannot be applied to, or the suggestions can be less helpful in, recently released programming problems.
However, this also indicates that the reliability can be increased as the number of solutions increases.

Second, it currently supports Python programming language only because of the dependency on the programming language in the normalization phase (e.g., tokenizer and formatter).
However, the fundamental idea of the proposed approach is versatile enough to be applied to any programming language if the normalization can be supported for the language, 

To the best of our knowledge, this work is the first investigation on deduplicating solution programs for suggesting common solution programs as references.
These limitations are expected to be mitigated in future work.

\section{Conclusion}
\label{sec:conclusion}

In this paper, we proposed an approach to deduplicate and rank solution programs in programming education, with the aim of suggesting common and helpful reference solutions to learners.
We performed experiments using a dataset of introductory programming problems from Aizu Online Judge to evaluate the effectiveness of the approach.

The experimental results showed that the approach significantly reduced the number of programs after deduplication compared to a baseline approach, with an average reduction of 60.20\%.
This means that users only need to refer to 39.80\% of the programs to grasp all the solution approaches on average.
Furthermore, we found that the top-10 ranked programs covered an average of 29.95\% of the programs.
This demonstrates that users can grasp nearly 30\% of the solution approaches by referring to only 10 programs.

The qualitative evaluation of the suggested programs also confirmed that they were mostly readable and could serve as helpful reference solutions.
However, we identified some limitations of the approach, such as the code format and the need for a sufficient number of solution programs for reliable suggestions.

In conclusion, our proposed approach shows promise in reducing the burden on learners to refer to too many solutions and motivating them to learn a variety of better approaches.
Future work includes incorporating the proposed approach into a real-world online judge system and evaluating its effectiveness in an actual educational setting.

\begin{acks}
This work was supported by the Japan Society for the Promotion of Science (JSPS) KAKENHI Grant Number JP23H03508.
\end{acks}

\bibliographystyle{ACM-Reference-Format}
\bibliography{main}

\end{document}